\documentclass[12pt]{article}
\font\titlefont=cmbx10 scaled \magstep4
\begin{document}

\begin{center}
{\titlefont The Classical Singularity Theorems \\ 
            and their Quantum Loopholes}
 
\vskip .3in
L.H. Ford\footnote{email: ford@cosmos.phy.tufts.edu} \\
\vskip .1in
Institute of Cosmology,
Department of Physics and Astronomy\\
Tufts University\\
Medford, Massachusetts 02155\\
\end{center}

\vskip .2in
\begin{abstract}
The singularity theorems of classical general relativity are briefly reviewed.
The extent to which their conclusions might still apply when quantum theory
is taken into account is discussed. There are two distinct quantum loopholes:
quantum violation of the classical energy conditions, and the presence of
quantum fluctuations of the spacetime geometry. The possible significance of 
each is discussed. 
\end{abstract}

\baselineskip=14pt

\section{Introduction: The Classical Singularity Theorems}

It has long been recognized that many solutions of Einstein's equations
contain curvature singularities, where the equations fail. There are two
cases of particular interest: the initial singularity in cosmological
models and the singularity in the interior of a black hole. The primary
example of the former is the Big Bang singularity at $t=0$ in a 
Friedman-Robertson-Walker model, whereas that of the latter is the
singularity at $r=0$ in the Schwarzschild solution. By the early 1960's, it
was recognized that both of these singularities posed a serious challenge
to classical general relativity. However, views differed as to whether
they are an artifact of the high degree of symmetry of the known examples,
or whether they are generic features that are to be expected even in solutions
with no symmetry. Among the proponents of the former view were Belinsky,
Khalatnikov, and Lifshitz \cite{BKL}, who attempted to represent the general
solution of a cosmological model near $t=0$ in a power series expansion.
Because lack of symmetry makes finding a generic exact solution a formidable
task, their aim was to learn as much as possible through approximate solutions.
A totally different approach was taken by Penrose, Hawking, and  others.
This was the development of {\it global techniques}. These techniques allow
one to prove, under certain assumptions, singularity theorems. These theorems
are now generally accepted as proving that singularities are indeed generic  
and not artifacts of symmetry. Here I will attempt to give only a very brief 
summary of global techniques. For more information, see the books by
Hawking and Ellis~\cite{HE} and by Wald~\cite{Wald}.

There are a variety of singularity theorems, but they typically
make four classes of assumptions:
\begin{itemize}
\item A classical spacetime obeying Einstein's equations. This simply says that
we are working in the framework of classical general relativity theory.
\item A stress tensor which satisfies an {\it energy condition}. Some 
restriction on the stress tensor is usually essential~\cite{BGV}, 
because every spacetime is
a solution of Einstein's equations with {\it some} stress tensor.
\item Some assumptions, such as the existence of a trapped 
surface, which specify the type of physical situation being discussed.
 These assumptions are also essential, as there are many nonsingular
exact solutions of the Einstein's equations, such as those which describe
static stars.  
\item An assumption concerning global behavior, such as an open universe
which is globally hyperbolic. 
\end{itemize}

Here are some examples of the energy conditions on the stress tensor 
$T^{\mu\nu}$ that might be assumed in the proof of a singularity theorem:
\begin{enumerate}

\item {\bf The strong energy condition.} $(T^{\mu\nu} 
-\frac{1}{2}g^{\mu\nu}\, T) u_\mu u_\nu \geq 0$, for all timelike vectors
$u^\mu$. Here $T = T^\mu_\mu$. In the  frame in which $T^{\mu\nu}$ is diagonal,
 this condition implies that the local energy density 
$\rho$ plus the sum of the local pressures $p^i$ is non-negative:
$\rho +\sum_i p^i \geq 0$, and that $\rho + p^i \geq 0$ for each $p^i$.
 This condition certainly holds for ordinary
forms of matter, although it can be violated by a classical massive scalar 
field.

\item {\bf The weak energy condition.} $T^{\mu\nu} u_\mu u_\nu \geq 0$, for all 
timelike vectors $u^\mu$. This condition requires that the local
energy density be non-negative in every observer's rest frame. 
 Again this seems to be a very
reasonable condition from the viewpoint of classical physics. 

\item {\bf The null energy condition.} $T^{\mu\nu} n_\mu n_\nu \geq 0$, for all
null vectors $n^\mu$. This condition is implicit in the weak energy condition.
That is, if we assume the weak energy condition, then the null energy condition
follows by continuity as $u^\mu$ approaches a null vector.

\item {\bf An averaged weak energy condition.} 
$\int T^{\mu\nu} u_\mu u_\nu d\tau \geq 0$, for all timelike geodesics, where
the integral is to be taken along either an entire geodesic with affine
parameter $\tau$, or a half-geodesic. These integral conditions are clearly 
weaker than the 
weak energy condition. It is now possible for the local energy density to be 
negative in some regions, so long as the integrated energy density is 
non-negative. 

\item {\bf An averaged null energy condition.} 
$\int T^{\mu\nu} n_\mu n_\nu d\lambda \geq 0$, for all null geodesics
(or half-geodesics), where
now $\lambda$ is the affine parameter. 

\end{enumerate}

A key result which is used to link the energy conditions to the properties
of spacetime is the Raychaudhuri equation for the expansion $\theta$ along
a bundle of timelike or null rays. It takes the form
\begin{equation}
\frac{d \theta}{d \tau} = -R^{\mu\nu} u_\mu u_\nu + 
({\rm other\,\, terms})\,,
\end{equation}
where $u^\mu$ is the tangent vector to the rays, 
$R^{\mu\nu}$ is the Ricci tensor, and the ``other terms'' are non-positive
for rotation-free geodesics. If the stress tensor satisfies the strong energy
condition, then the Einstein equations,
\begin{equation}
R^{\mu\nu} = 8\pi (T^{\mu\nu} -\frac{1}{2}g^{\mu\nu}\, T) \, ,
\end{equation}
imply that
\begin{equation}
\frac{d \theta}{d \tau} < 0 \, .
\end{equation}
This is the condition that the bundle of rays is being focused by the
gravitational field, and is consistent with our intuition that gravity
is attractive.

The basic strategy to prove a singularity theorem is essentially the following:
one assumes an energy condition and infers the presence of focusing.
This is then combined with additional assumptions to infer the existence
of extremal length geodesics. An example would be a timelike geodesic which 
ends in a finite proper time. Finally, one infers the existence of a
singularity by the existence of non-extendible geodesics. The basic idea is that
if spacetime is non-singular, all geodesics should be extendible over an
infinite range of affine parameter. 

The first theorem to be proven was Penrose's
theorem \cite{P65}, which implies that singularities must arise when a black 
hole is formed by gravitational collapse. In addition to some technical 
assumptions, the proof of this theorem relies upon an energy condition and
on the assumption of the existence of a trapped surface. Such a surface arises 
when the gravitational field of the collapsing body becomes so strong that
outgoing light rays are pulled back toward the body. Penrose's original proof
assumed the weak energy condition, but later authors~\cite{T78,G81,B87,R88}
 were able to provide proofs of this and other theorems that assume only an 
averaged  energy condition. The essence of 
the theorem is that so long as either of these energy conditions is obeyed,
once gravitational collapse proceeds to the point of formation of a trapped
surface, then the formation of a singularity is inevitable. Penrose \cite{P02} 
has recently suggested that a variant of this theorem might rule out the 
existence of compact extra dimensions of the sort postulated in Kaluza-Klein 
theories. The basic idea is that the wrapping of light rays around the compact 
dimensions would create an effect analogous to the trapped surface in
gravitational collapse.

Some general comments about singularity theorems are now in order. The global
techniques used in their proofs are very general and powerful. For example,
there is no need to assume any symmetry and no need to try to solve the
Einstein equations. On the other hand, the theorems say very little about the
nature of the singularity. Penrose's theorem proves the existence of a
non-extendible geodesic. One suspects that this must be due to the formation 
of a curvature singularity, as happens in the spherically symmetric case,
but there is no proof of this.  The drawback of the global methods
is that they rely upon indirect arguments and proof by contradiction. 
This makes them perhaps less robust against loopholes in their assumptions, 
so it is necessary to examine these assumptions critically, especially 
in the light of quantum effects.

\section{Quantum Loophole \# 1: Violation of the Energy Conditions}

It is well-known that quantum effects can indeed violate classical energy
conditions, such as the weak energy condition. In particular, quantum effects
can give rise to negative local energy densities. An example of this
is the Casimir effect: the electromagnetic vacuum state between a pair of 
perfectly conducting plates has an energy density of
\begin{equation}
\rho = - \frac{\pi^2}{720 L^4} \, ,
\end{equation}
where $L$ is the plate separation and units in which $\hbar = c =1$ are used.
This violates both the weak and the averaged weak energy conditions, as an
observer between the plates at rest observes a constant negative energy density.
Interestingly, the averaged null energy condition is not clearly violated
in this case. The only null rays which avoid hitting the plates (and hence
their presumably large positive energy density) are those which are parallel
to the plates. In this case, $T^{\mu\nu} n_\mu n_\nu = 0$, so the averaged 
null energy condition is marginally satisfied. One might wonder if the
violation of the weak energy condition by the Casimir effect is an artifact
of the assumption of perfectly reflecting boundary conditions. It has
recently been shown~\cite{VF02} that more realistic plates with finite, but
sufficiently high, reflectivity can also produce negative local energy density.
In all cases, there is an inverse relation between the size of the negative
energy region (the plate separation) and the magnitude of the negative energy
density. 

A second way that quantum effects can create negative energy density is 
through quantum coherence effects. One can construct quantum states in
a quantum field theory in which the local energy density is negative. The
simplest example of this is a quantum state for a bosonic field which is 
superposition of the vacuum and of a two particle state for a particular mode:
\begin{equation}
|\psi \rangle = N (|0 \rangle + \epsilon |2 \rangle) \,,
\end{equation}
where $N$ is a normalization factor and $\epsilon$ is the relative amplitude 
to measure two particles rather than no particles in the state. In Minkowski
spacetime, the local energy density is the expectation value of the normal
ordered stress tensor operator, $:T_{tt}:$,
\begin{equation}
\rho = \langle \psi| :T_{tt}: |\psi \rangle =
 N^2 [2 Re(\epsilon \langle 0| :T_{tt}: |2 \rangle 
 + |\epsilon|^2 \langle 2| :T_{tt}: |2 \rangle] \,.
\end{equation}
The only other piece of information that we need is that in general
$\langle 0| :T_{tt}: |2 \rangle \not= 0$. If we take $|\epsilon|$ sufficiently
small, then the $|\epsilon|^2$ term in $\rho$ can be neglected, and we
can then choose the phase of $\epsilon$ so as to have $\rho < 0$ at a selected 
spacetime point. This state is essentially a limit of a squeezed vacuum
state. 

Although the local energy density in states such as that described above
can be made arbitrarily negative at a given spacetime point, one finds that
there are two important restrictions on the negative energy density, at
least for free fields in Minkowski spacetime. The first is that the total
energy must be non-negative:
\begin{equation}
\int \rho\, d^3x \geq 0 \,.
\end{equation}
The second is that the energy density integrated along a geodesic
 observer's worldline 
with a sampling function $f(\tau)$ must obey a ``quantum inequality'' of
the form~\cite{F91,FR95,FR97,FE}
\begin{equation}
\int_{-\infty}^\infty \rho(\tau)\,f(\tau)\, d\tau \geq - \frac{c}{\tau_0^4} \,,
                                                   \label{eq:QI}
\end{equation}
where $\tau_0$ is the characteristic width of $f(\tau)$ and $c$ is a
dimensionless constant, which is typically somewhat less than unity.
The physical content of these inequalities is that there is an inverse relation
between the magnitude of negative energy density, and its duration. An observer
who sees a negative energy density of magnitude $|\rho_m|$ will not see
it persist for a time longer than about $|\rho_m|^{-1/4}$. This restriction
greatly limits what one can do with quantum negative energy. Macroscopic
violations of the second law of thermodynamics, which would occur with
unlimited negative energy, seem to be ruled out~\cite{F78}, for example.

Quantum inequalities have been proven under a variety of conditions to hold
in curved spacetime~\cite{FLAN,PF971,PFGQI,Fewster}, as well as in flat 
spacetime. In particular, if the sampling time $\tau_0$ is small compared to $r$, 
the local radius of curvature, 
then the flat space form, Eq.~(\ref{eq:QI}), is approximately valid in curved 
spacetime as well. The inequalities basically say that the local energy
density cannot be vastly more negative than about $-1/r^4$.
 This fact has been used to place severe restrictions on
some of the more exotic gravitational phenomena which negative energy might 
allow, such as traversable wormholes~\cite{FRWH}, 
or ``warp drive'' spacetimes~\cite{PFWD}. 

The key question remains: can quantum violations of the energy conditions
avoid the singularities of classical relativity? In at least some cases, the
answer is yes. An example of this was given many years ago by Parker and 
Fulling~\cite{PF73}, who constructed a non-singular cosmology using
quantum coherence effects to avoid an initial singularity. These authors
explicitly constructed a quantum state which violates the strong energy
condition and in which the universe would bounce at a finite curvature,
rather than passing through a curvature singularity. Furthermore, the
bounce can be at a scale far away from the Planck scale. This example shows
that quantum effects can avoid an initial cosmological singularity, but
leaves open the question of whether the singularity is necessarily avoided
by quantum processes.

The case of the black hole singularity is technically more difficult to
study, and no explicit construction analogous to the Parker-Fulling example in
cosmology has been given. However, several authors have discussed the form 
which non-singular black holes might take. Frolov, Markov, and 
Mukhanov~\cite{FMM}, for example, have discussed the possibility that the
Schwarzschild geometry might make a transition to a deSitter spacetime before
the $r=0$ singularity is reached. 

Most of the work on quantum singularity avoidance has been in the
context of a semiclassical theory, where matter fields are quantized but 
gravity is not. This theory should break down before the Planck scale
is reached, at which point one would need a more complete theory. It is
not clear that one can get generic singularity avoidance in this theory
far away from the Planck scale. One can give a simple argument for this:
In Planck units, quantum stress tensors typically have a magnitude of 
the order of $\langle T^{\mu\nu} \rangle \sim 1/r^4$, whereas the
Einstein tensor is of order $G^{\mu\nu} \sim 1/r^2$, where $r$ is the 
local radius of curvature. The backreaction of the quantum field on
the spacetime geometry is large when $\langle T^{\mu\nu} \rangle
\approx G^{\mu\nu}$, which is when $r \approx 1$, that is, at the Planck
scale. Of course, this argument does not always hold, as the Parker-Fulling 
example shows. However, the reason that  Parker and Fulling were able
to get a bounce well away from the Planck scale is twofold: Their example
requires negative pressure, but not negative energy density (violation
of the strong but not the weak energy condition), and their model contains
a massive field, introducing a new length scale. Thus, in their example,
the violation of the appropriate energy condition is not characterized
by $1/r^4$. However, the quantum inequalities seem to suggest that one
cannot get such large violations of the weak energy condition, and that
local negative energy densities in curved spacetime are likely to be
of order $-1/r^4$. 

 It should be noted that it is possible to violate energy
conditions at the classical level with nonminimally coupled scalar fields,
and this fact has been used by Saa, {\it et al}~\cite{Saa} to construct
nonsingular cosmologies with such fields as the matter source. Thus if
there are such nonminimal fields in nature, all of the discussion of
quantum violation of the energy conditions may be moot.

\section{Quantum Loophole \# 2: Quantum Fluctuations of Spacetime Geometry}

There is another, very different, loophole in the classical global analysis 
which is posed by quantum effects. This is the presence of fluctuations
of the spacetime geometry. These fluctuations have been discussed in recent 
years by many authors. See Refs.~\cite{JR95,P01,MV00,S00,HV02,F95,KF93} for 
a partial list.
 There are two sources for these fluctuations.
One is the fluctuations which arise when the gravitational field itself
is treated as a quantum field, which might be called the ``active'' 
fluctuations.
The second source is quantum fluctuations of the stress tensor of a
quantized matter field. Even in a theory in which gravity itself is not 
described quantum mechanically, fluctuations of the local energy density
will drive fluctuations of the gravitational field. The presence of these
fluctuations means that the assumption of a classical spacetime obeying 
Einstein's equations is not strictly valid. Light rays in general no longer
precisely focus as they would on a fixed classical spacetime.

We can quantify this by treating the Raychaudhuri equation, Eq.~(1), as
a Langevin equation, with a fluctuating Ricci tensor term. This can be
done regardless of the source of the fluctuations. Then one can find the 
dispersion in $\theta$ as an integral of the Ricci tensor correlation
function:
\begin{equation}
 \langle \theta^2 \rangle - \langle \theta \rangle^2
= \int d\lambda \int d\lambda' \, u^\mu u^\nu u^\alpha u^\beta
[\langle R_{\mu\nu}(\lambda) R_{\alpha\beta}(\lambda')\rangle
- \langle R_{\mu\nu}(\lambda)\rangle \langle R_{\alpha\beta}(\lambda')\rangle].
\end{equation}

In many contexts, the quantum fluctuations of the metric are expected to
be a very small effect. For example, in the collapse of a star to form a
black hole, the root-mean-square fluctuations of the Ricci  tensor
are likely to be very small compared to the classical Ricci  tensor of
the collapse spacetime, at least until very close to the singularity.
The problem for global techniques, is the indirect nature of the arguments,
such as the reliance on exact focusing and on proof by contradiction.  
Thus, even if the
conclusions of the singularity theorems are still correct, in the presence
of fluctuations the proofs are not strictly valid.

\section{Summary}

The classical singularity theorems are based on very powerful indirect 
arguments which show that black hole and cosmological singularities are
generic in classical general relativity, meaning that the theory breaks down.
This suggests that a way to avoid this problem must be found in a new theory,
such as one incorporating quantum effects. Quantum violations of the classical
energy conditions certainly open this possibility. However, such violations
tend to occur on short distance scales, or at high curvatures. Furthermore,
one may need to go beyond a semiclassical theory to a more complete quantum
theory of gravity in order to understand how quantum theory avoids 
singularities.

The presence of fluctuations also poses a challenge for global techniques,
with their reliance on exact focusing.
Yet it is hard to see why a very small fluctuation would qualitatively
change the behavior of a gravitational field. Thus, it may well be that 
small quantum fluctuations do not prevent large curvatures from being reached
in the early universe or inside a black hole. However, to prove this one
will need new methods. 

\vspace{0.5cm}

{\bf Acknowledgements:} I would like to thank Arvind Borde and Tom Roman for 
helpful comments on the manuscript. This work was supported in part by the 
National Science Foundation under Grant PHY-9800965.

\end{document}